\documentclass[runningheads]{llncs}
\usepackage[T1]{fontenc}
\usepackage{lipsum}
\usepackage{graphicx}
\usepackage{amsmath,amssymb,amsfonts}
\usepackage{algorithmic}
\usepackage{algorithm}
\usepackage{textcomp}
\usepackage{stfloats}
\usepackage{mathtools}

\DeclarePairedDelimiter\floor{\lfloor}{\rfloor}

\usepackage{mathtools}
\usepackage{graphicx}
\usepackage{textcomp}
\usepackage{xcolor}
\usepackage{subcaption}
\usepackage{caption}

\begin{document}
\title{LogSHIELD: A Graph-based Real-time Anomaly Detection Framework using Frequency Analysis}

%
%
\author{Krishna Chandra Roy\inst{1}\orcidID{0000-0001-9388-8042} \and
Qian Chen\inst{2}\orcidID{0000-0002-0130-5901}}
\authorrunning{K. Roy et al.}
%
\institute{New Mexico Institute of Mining and Technology, Socorro NM 87801, USA \and
University of Texas at San Antonio, San Antonio TX 78249, USA
\email{krishna.roy@nmt.edu, guenevereqian.chen@utsa.edu}}
\maketitle              
\begin{abstract}
Anomaly-based cyber threat detection using deep learning is on a constant growth in popularity for novel cyber-attack detection and forensics. A robust, efficient, and real-time threat detector in a large-scale operational enterprise network requires high accuracy, high fidelity, and a high throughput model to detect malicious activities. Traditional anomaly-based detection models, however, suffer from high computational overhead and low detection accuracy, making them unsuitable for real-time threat detection. In this work, we propose \textbf{LogSHIELD}, a highly effective graph-based anomaly detection model in host data. We present a real-time threat detection approach using frequency-domain analysis of provenance graphs. To demonstrate the significance of graph-based frequency analysis we proposed two approaches. Approach-I uses a Graph Neural Network (GNN) LogGNN and approach-II performs frequency domain analysis on graph node samples for graph embedding. Both approaches use a statistical clustering algorithm for anomaly detection. The proposed models are evaluated using a large host log dataset consisting of 774M benign logs and 375K malware logs. LogSHIELD explores the provenance graph to extract contextual and causal relationships among logs, exposing abnormal activities. It can detect stealthy and sophisticated attacks with over 98\% average AUC and F1 scores. It significantly improves throughput, achieves an average detection latency of 0.13 seconds, and outperforms state-of-the-art models in detection time.



\end{abstract}

\section{Introduction}
\label{sec:introduction}

Anomaly-based intrusion detection approaches are indispensable and widely adopted for zero-day attack detection, but they suffer from a high false-alarm rate compared to signature and rule-based detectors\cite{zhang2019robust}. Although traditional signature and rule-based approaches achieve higher detection accuracy, they fail to detect sophisticated and novel attack campaigns as these detectors depend on malware signatures. Anomaly-based detectors analyze flow-based data such as NetFlow, leveraging a set of flow fields to model anomaly behavior patterns \cite{lo2021graphsage}. However, the flow sequence analysis models suffer from modeling the long-term dependency of the network activity which limits the detection capability. To this end, the provenance graph of system logs provides a better representation of systems and network activities describing the information flow between system processes for anomaly detection\cite{han2020unicorn}. Graph Neural Networks (GNNs) can capture the rich node semantics of provenance graphs by leveraging graph neighborhoods. Therefore, Machine learning/Deep learning detectors show promising anomaly detection performance utilizing GNN-based provenance graph semantics \cite{roy2024graphch,zheng2018realtime}.

As the size of provenance graphs grows, graph-based detectors become computationally expensive. The high computational overhead and increased processing time of deep learning algorithms cause the models to struggle with handling a high volume of audit data. Therefore, deep learning-based computation-intensive models can be deployed for offline detection but are challenging to use for real-time detection\cite{du2019lifelong,zhu2020you}.

Frequency domain analysis (FDA) of audit data is an emerging technique for reducing the computational overload of graph-based detection models. FDA leverages the transformation of graphs into the frequency domain to identify anomalous behaviors \cite{fu2021realtime}. To leverage a graph representation of audit data and design a real-time anomaly detection model we propose LogSHIELD, a host data-driven provenance graph analysis framework exploiting FDA. LogSHIELD utilizes the causal, sequential, and logical relationship of system logs to construct a provenance graph. To demonstrate the trade-off between detection accuracy and detection time, we propose two graph analysis approaches. Approach-I is LogSHIELD with LogGNN, a GNN-based approach for embedding graph nodes and identifying abnormal sequences. Approach-II is LogSHIELD with FDA\cite{fu2021realtime}, a Frequency Domain Analysis (FDA) approach for embedding graph nodes. Upon graph embedding both approaches apply a statistical clustering algorithm using the cosine similarity of the node embedding vectors for anomaly detection.      

Approach-I, LogSHIELD with LogGNN analyzes the log provenance graph and generates effective log embedding. However, one major drawback is embedding time and detection latency due to the complex embedding network. To handle the higher detection latency of the deep embedding model, approach-II performs frequency domain analysis on the graph. A bidirectional Long Short-Term Memory (BiLSTM) model in LogGNN is replaced with a frequency analysis FDA module. FDA extracts high-level features from the log provenance graph without utilizing computation-intensive deep learning models which is the key advantage of approach-II. LogSHIELD with FDA takes significantly less detection time and achieves higher throughput compromising a small fraction of detection accuracy.


The design and implementation of LogSHIELD analyzing host logs pose the following challenges. The first challenge is constructing a provenance graph using redundant and concurrent host logs. Many redundant and system-level noisy logs are generated which causes the graph to have thousands of redundant nodes and edges. To solve this challenge we pre-process the raw logs and statistically filter the redundant and noisy logs. The log messages contain rich information as event fields such as {\tt EventID, ProcessName, ProcessID, Timestamp}, etc. Event fields are not uniform among all event logs and differ with system environments. For example, the benign host dataset contains $1200$ unique fields. Therefore, the second challenge is to down-sample the event fields and select the right fields for content encoding. To solve this challenge we analyzed the 2TB of host data and found the unique and relevant log fields representing log semantics. The third challenge is a needle in a haystack. Hundreds of system logs are generated within seconds depending on the running processes, which makes detection challenging for real-time applications. Moreover, malicious logs are a very small fraction of the benign system logs which makes detection even harder. We solve this challenge by designing a robust and scalable analysis and detection framework. LogSHIELD constructs a provenance graph utilizing the activity logs of the host computer creating causal, sequential, and logical connections. LogSHIELD approach-I utilizes the LogGNN embedding algorithm to learn the underlying representation of the log graph. Random walk-based node sampling in LogGNN provides high throughput in analyzing the logs in large enterprise networks with thousands of users.

The main contributions of this paper are as follows:
\begin{itemize}
    \item We propose a real-time anomaly detection framework LogSHIELD with FDA to achieve high throughput detection for real-time applications.
    \item We also propose LogSHIELD with LogGNN, a robust and high-fidelity anomaly detection framework using Graph Neural Network LogGNN to demonstrate the trade-off between detection accuracy and detection time.
    \item A major challenge in model evaluation is the insufficient real-world host log dataset. To this end, we present a real-world benign and malware dataset. The benign dataset contains more than 774M of host logs for 35 Windows machines and the malware dataset contains logs for 140 malware samples.
    \item LogSHIELD evaluated with the two real-world log datasets. The results validate the effectiveness of LogSHIELD for real-time and high-accuracy anomaly detection.
    \item An in-depth ablation study is performed for both the proposed approaches to evaluate functional modules and demonstrate the trade-off between detection accuracy and detection time. 
\end{itemize}

The remainder of the paper is organized as follows. Section~\ref{sec:relatedwork} presents the related works in real-time and graph-based anomaly detection. In Section~\ref{sec:framework} we introduce the LogSHIELD framework and describe the design approaches. Section~\ref{sec:experimental_evaluation} presents the experimental design, results, and result evaluations. Finally, Section~\ref{sec:conclusion} concludes the study.

\section{Related Work}
\label{sec:relatedwork}
This section presents existing research on graph-based anomaly detection, focusing on approaches that utilize deep learning, and machine learning algorithms. 

\subsection{Real-time Anomaly Detection}
Early detection and prevention of threat actors can mitigate the profound losses incurred by cyber threats. However, sophisticated systems and stealthy attacks require real-time detection. Many researchers working toward reducing the detection time and achieving real-time anomaly detection \cite{irshad2021trace,han2022kernel,wang2022lightlog,wu2022paradise,han2020unicorn,fu2021realtime,panda2022host,yang2022enhanced}. 

In \cite{wu2022paradise}, Wu et al. proposed "Paradise," a real-time, generalized, and distributed provenance-based intrusion detection system. Paradise extracts process feature vectors in a separate environment at the system log level and stores them in high-efficiency memory databases. During the detection phase, it calculates provenance-based dependencies for intrusion detection. Another graph-based real-time outlier detection technique is proposed in \cite{panda2022host}, which learns all possible process relation semantics. The authors claim that a generalized process relation graph (GPRG) enables the detection of any outlier program with 96\% accuracy at any time instance. In \cite{irshad2021trace}, Irshad et al. proposed TRACE, a scalable, real-time, enterprise-wide provenance tracking system for APT detection. It uses static analysis techniques to identify unit dependencies and strategies to generate concise provenance graphs, evaluated during five red-team engagements. In \cite{wang2022lightlog}, Wang et al. proposed LightLog, a lightweight temporal convolutional network (TCN) for log anomaly detection in edge devices. LightLog uses word2vec to generate low-dimensional semantic vectors from logs and TCN for detection, achieving real-time processing and detection. Static log stream feature analysis approaches are also adopted by many researchers for real-time anomaly detection. In \cite{fu2021realtime}, Fu et al. proposed "Whisper," a real-time robust malicious network traffic detection method using frequency analysis. The authors experimented with 42 types of attacks and demonstrated that Whisper achieves high throughput with 99\% AUC accuracy. While our proposed framework LogSHIELD with FDA is inspired by Whisper, there are notable differences. Unlike Whisper, which uses network traffic features and performs frequency analysis for detection, LogSHIELD constructs a provenance graph from host logs, performs RWR node sampling, and then applies frequency analysis to the sampled node contents.

\subsection{Graph-based Anomaly Detection}

Graph representations of multidimensional and heterogeneous data are used in modeling system behavior, thereby detecting abnormalities and threats within a system. Many research works have applied graph-based analysis techniques to threat detection. In \cite{liu2019log2vec}, Liu et al. proposed a graph-based user behavior modeling approach for insider threat detection using heterogeneous graph embedding. Log2vec uses a set of rules to construct the graph and represents each log entry as a low-dimensional vector using a random walk. In \cite{wang2018learning}, Wang et al. introduced an approach for automatic insider threat detection that utilizes not only self-anomalous behaviors of an employee but also anomalies relative to other employees with similar job roles. It infers the correlation graph among the organization’s employees and identifies potential threats using graph signal processing. A graphical analysis and anomaly detection-based approach was proposed in \cite{gamachchi2018graph}, introducing a hybrid framework consisting of a graphical processing unit and an anomaly detection unit to isolate anomalous users from heterogeneous data (logon/logoff, email logs, HTTP logs, etc.). It also incorporated psychometric data. In another paper, they used an attributed graph clustering technique and an outlier ranking mechanism for threat detection \cite{gamachchi2017insider}. In \cite{le2021training}, graph-based semi-supervised machine learning methods, such as label propagation and label spreading, were used for insider threat detection in the CERT insider threat test dataset. Many researchers utilize provenance graphs for anomaly detection. In \cite{han2020unicorn}, Han et al. proposed UNICORN, a provenance graph-based advanced persistent threat detection model that uses a histogram and hashing algorithm for graph feature extraction.

\begin{figure*}[h]
  \centering
  \begin{minipage}[b]{0.9\textwidth}
    \includegraphics[width=\textwidth]{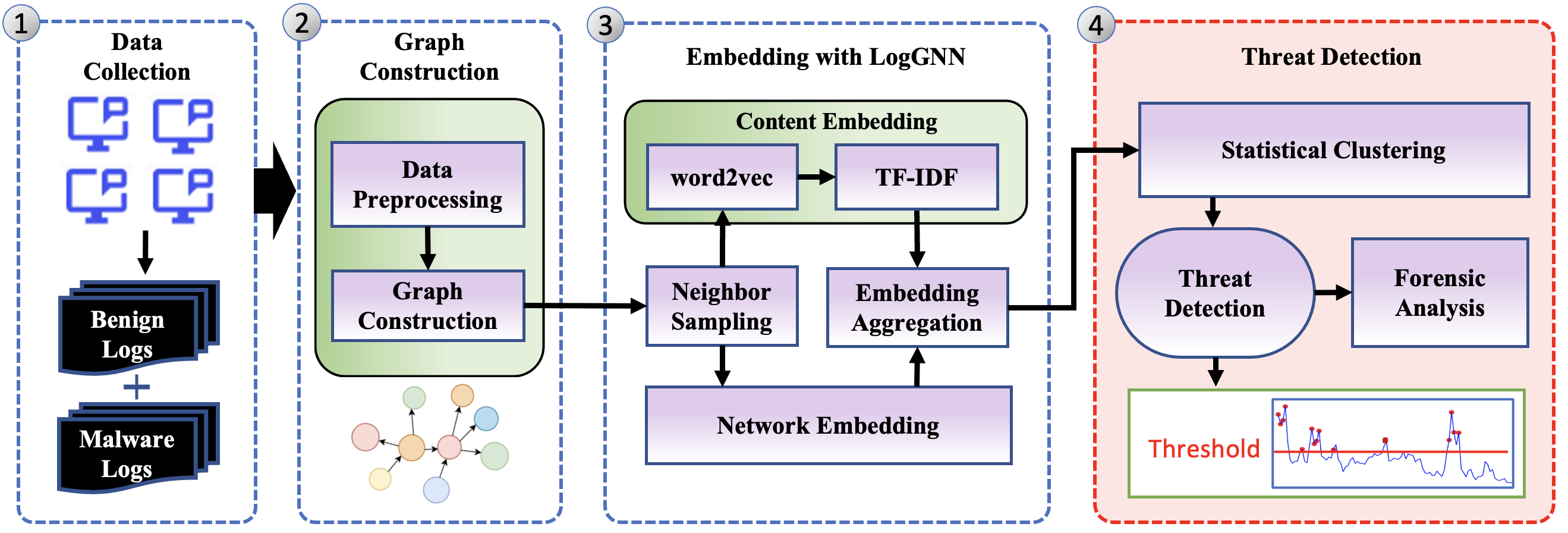}
    \caption{Approach-I: System architecture of LogSHIELD with LogGNN graph embedding.}
    \label{fig:architecture}
  \end{minipage}
\end{figure*}

\section{Proposed L\MakeLowercase{og}SHIELD Framework}
\label{sec:framework}
We introduce two approaches for graph high-level feature extraction and a statistical clustering algorithm for anomaly detection.

\subsection{\textbf{Approach-I:} LogSHIELD with LogGNN}
Approach-I consists of four main steps: log data collection, graph construction, graph embedding with LogGNN, and anomaly detection. A constructional diagram of LogSHIELD with LogGNN is presented in Figure~\ref{fig:architecture}. LogGNN learns the representation and dependency of the nodes in the provenance graph and obtains a semantic embedding vector for each of the nodes in the graph.

\subsubsection{Log Preprocessing}
The logs in the benign and malware datasets are collected using Windows Logging Service (WLS) in standard JSON format. For the benign dataset, we recruit $35$ legitimate users from a large enterprise network and collect their working PCs' host logs over $90$ days upon approval from the Institutional Review Board (IRB) as human subjects are involved. For the malware dataset, we collected logs for 140 latest malware binaries deployed manually in a testbed with two target machines running Win10 OS. These two physical machines’ configuration is the same as the configuration of benign participants’ computers.

\begin{figure}
\centering
    \includegraphics[width=0.5\textwidth]{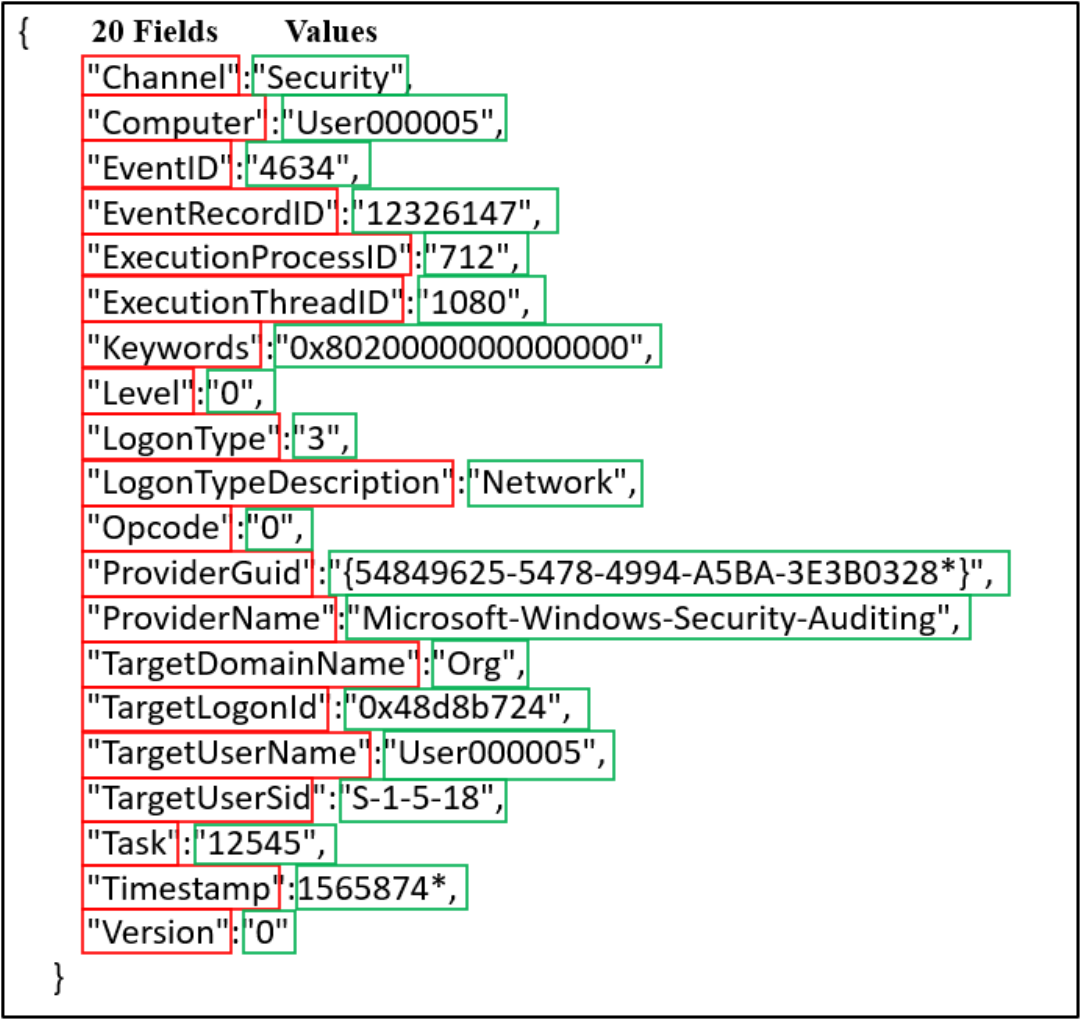}
    \caption{A host log example for {\tt EventID 4634}. Red boxes are event fields and green boxes are values of the corresponding fields.}
    \label{fig:event_example}
\end{figure}

A total of around $774$ million in event logs were collected where each event has its relevant field set with corresponding field values. We perform a data cleaning step before constructing a graph to filter the noisy and redundant system logs. We analyze the dataset and down-sample to 5 event fields {\tt EventID}, {\tt ProcessName}, {\tt BaseFileName}, {\tt LogonType}, {\tt ParentProcessName}, out of $1200$ unique fields as the content of the logs. These two steps of pre-processing reduce the volume of the host data to one-fifth of the 2TB raw data size which significantly reduces the log embedding time in LogGNN. Figure~\ref{fig:event_example} presents an example of a host log with EventID 4634, a logoff event.  

All events in the processed host logging dataset contain the  "EventID" field, which represents the processes triggered by users or computer programs. Hundreds of unique "EventID" values have been collected, and the number of unique "EventID" values varies by user's activity. For example, a sequence of five events has EventID values $4624$ $\rightarrow$ $4672$ $\rightarrow$ $4798$ $\rightarrow$ $4798$ $\rightarrow$ $4634$, meaning that (1) the user successfully logon ($EventID=4624$); (2) special privileges are assigned to new logon ($EventID=4672$); (3) a user's local group membership is enumerated by "chrome.exe," the Google Chrome application ($EventID=4798$); (4) the user's local group membership is enumerated by  "svchost.exe, " a Windows system process to host from one to many services ($EventID=4798$); and (5) the user is logged off ($EventID=4634$). A detailed description of host log collection can be found in \cite{acquesta2019detailed}.

\subsubsection{Graph Construction}
\label{subsubsec:graph_const}
To address system anomaly detection as a graph analysis problem, we construct a provenance graph from daily host logs within an enterprise network, capturing the interrelationship of log entries. This is step 2 in Approach-I as shown in Figure~\ref{fig:architecture}. The log provenance graph (LPG) focuses on three types of relationships: causal, sequential, and logical, to learn the contextual representation of event logs. For sequential relationships, we consider the timestamps of log entries to resolve out-of-order issues caused by log arrival delays. Causal relationships among processes (EventID 4688, 4689) are established by creating directional connections between parent and child processes using the fields ProcessName, BaseFileName, and ParentProcessName. Additionally, LogSHIELD builds logical relationships by connecting log entries with the same timestamp. LogSHIELD constructs the graph using five key event fields: EventID, ProcessName, BaseFileName, LogonType, and ParentProcessName. This approach ensures the graph construction follows these specific rules.

\begin{itemize} 
    \item Rule 1: All log entries from the same user are connected sequentially based on their timestamps, forming the main chain of triggered logs.
    \item Rule 2: Log entries from the same user that are related by a parent-child process are connected with a directional link.
    \item Rule 3: Log entries from the same user with the same timestamp (concurrent logs) are all connected to the last log of the chain with a distinct timestamp. 
    \item Rule 4: Log entries from the same user with the same timestamp maintain an internal connection between them.
\end{itemize}

\begin{figure}
\centering
    \includegraphics[width=0.6\textwidth]{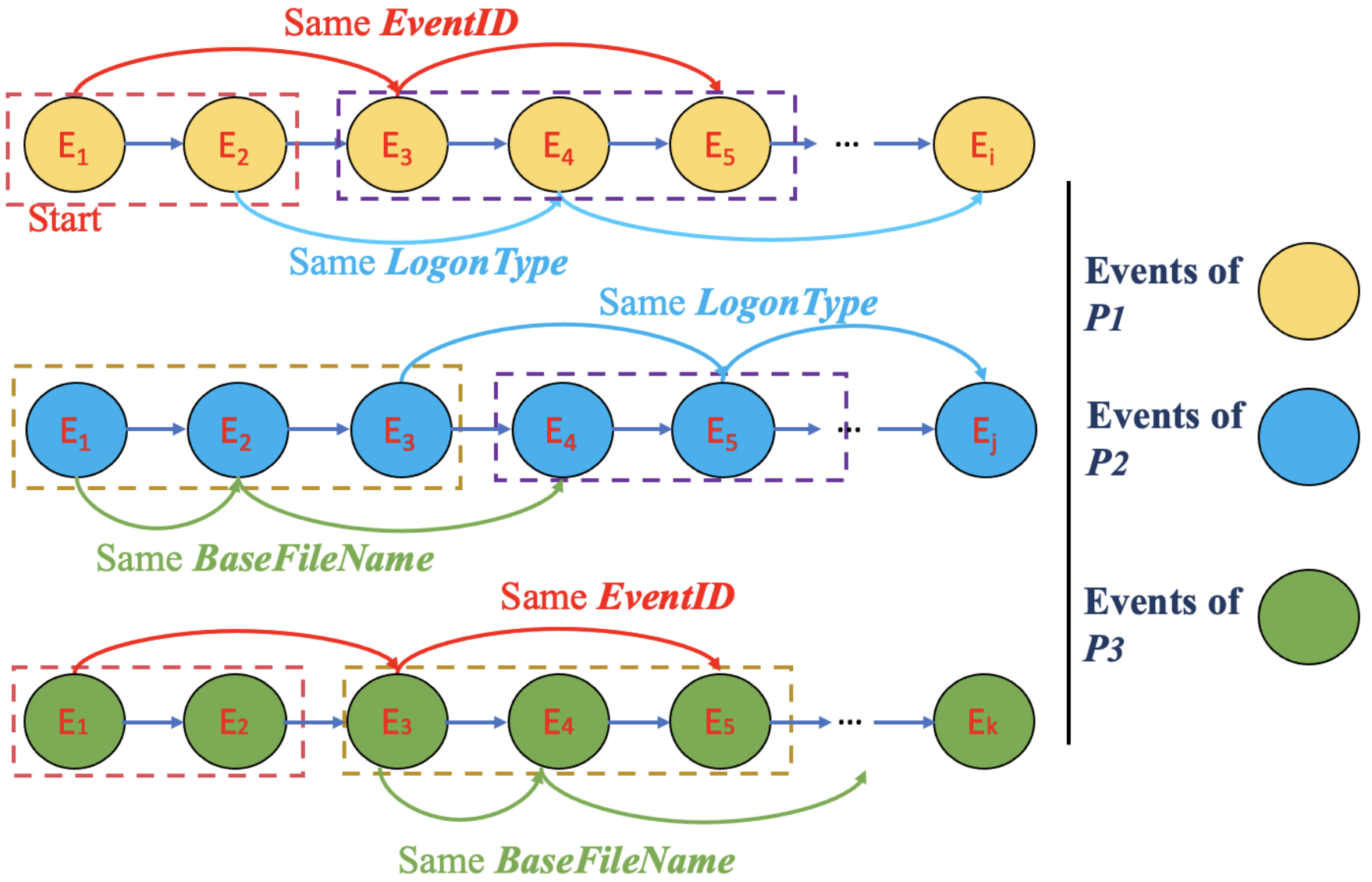}
    \caption{A provenance graph constructed using a small fraction of daily benign logs of User P1, P2, and P3.}
    \label{fig:prov_graph}
\end{figure}

Figure~\ref{fig:prov_graph} presents an example of a provenance graph constructed using a small fraction of daily benign logs of 3 randomly selected users P1, P2, and P3. 

\subsubsection{Log Neighbor Sampling}
\label{subsubsection:neigh_sampling}
This is the beginning of step 3 in approach-I. The neighbors of a node significantly contribute to its semantics in representation learning. Therefore, in the first step of graph embedding, it samples representative neighbor nodes, performs encoding, and aggregates them to find the embedding of the target node. Many node neighbor sampling algorithms have been adopted recently including random walk sampling \cite{zhang2019heterogeneous}, direct neighbor sampling with different order\cite{wang2019heterogeneous}. Direct neighbors can differ in number leading to insufficient representation of a node and an inability to capture diverse node types in the graph. Random walk helps to capture the heterogeneous node by sampling the n-hop neighbors.

Random Walk is one of the most popular neighbor sampling algorithms. In this work, we used a modified random walk with a restart strategy. It generates paths traversed by walkers starting from node $v \in V$. The walker travels through n-hop neighbors of the current node and returns back until a fixed number of nodes are sampled which is walk length, the walker traverses through the neighbors. We sample $\mathcal{S}(v)$ neighbor nodes of the node $v$.

\begin{figure}[h]
  \centering
    \includegraphics[width=0.5\textwidth]{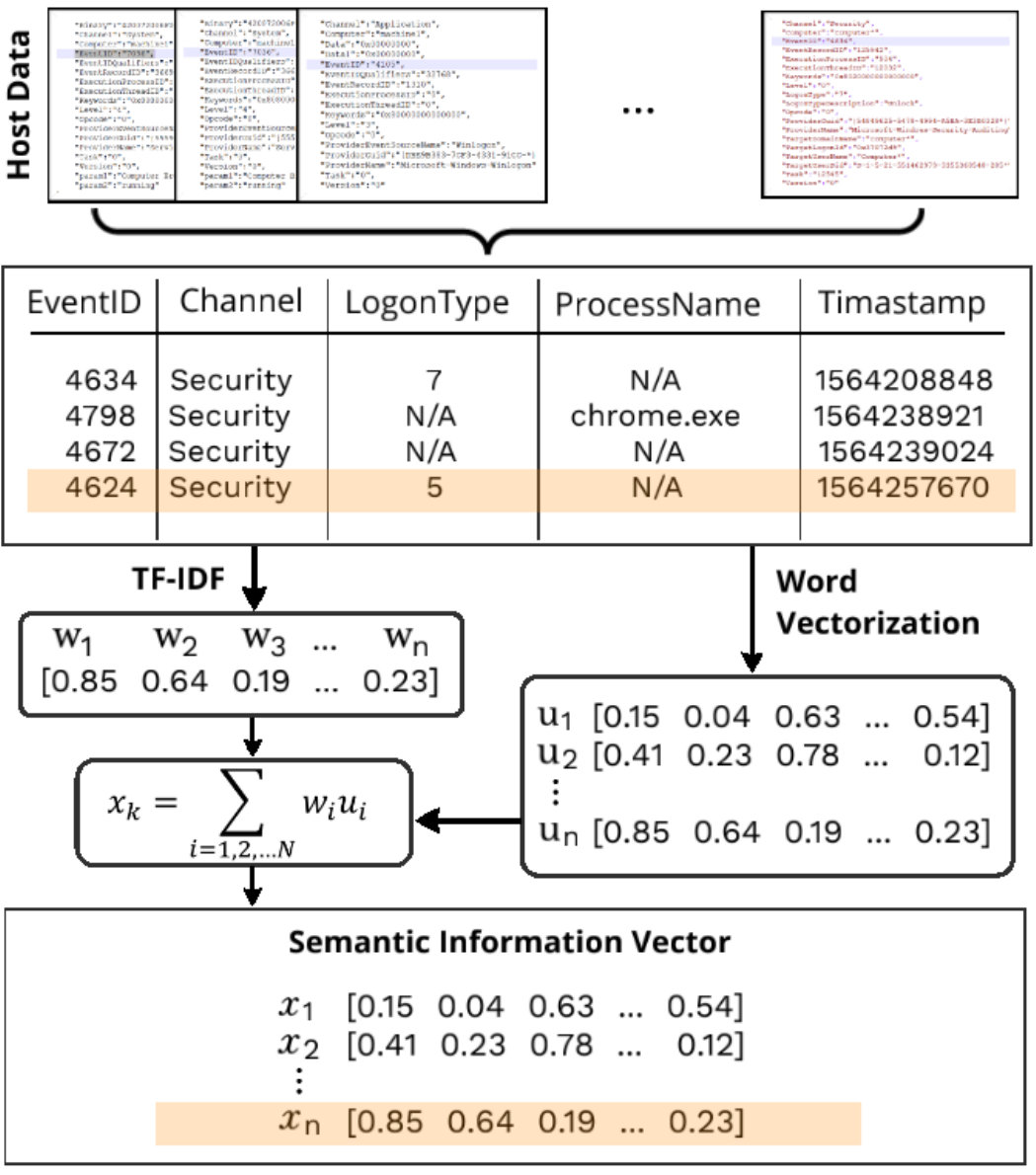}
    \caption{Log parsing workflow for approach-I. A random set of log event fields is shown in the parsing workflow.  }
    \label{fig:parsing}
\end{figure}

\subsubsection{Log and Network Encoding}
\label{subsubsec:loggnn_embedding}
The next step of graph embedding is encoding heterogeneous contents $C_v$, which are attributes or texts of the nodes from $v \in V$. LogGNN uses word2vec\cite{church2017word2vec} model for encoding log fields. Node content embedding is performed using a log parsing algorithm in \cite{roy2020deepran}. The details of the encoding text content of event nodes are presented in this section. In a similar fashion to  Natural Language Processing (NPL) that converts a group of sentences into tokens, the events in the host log are represented to tokens $T=[t_1, t_2,\cdots,t_n]$ as presented in Figure~\ref{fig:event_example}. In a host log file, the fields of the dictionary structure (e.g., BaseFileName, EventID) are the tokens, and the number of unique tokens selected from the event fields is $n$. The log file is converted to a sequence of event vectors $\mathcal{S}(v)=[ V_1, V_2,\cdots, V_K ]$. Where $K$ represents the $K-th$ node. After that, event vectors are represented as  $V_{i}= [v_1, v_2, \cdots,v_n]$. The next step of the parsing algorithm is word vectorization. We use FastText\cite{joulin2016fasttext}, an extension to the Word2Vec model for word vectorization. Each token parameter value $v_j$ ($j\in [1, n]$) of host events converted to a $e$-dimension vector $u_j$ (i.e., $v_j \rightarrow u_j=[a_1, a_2,\cdots,a_e]$). Vector $V_{i}$ is transposed as $V_{i}^{(T)}$, where $i\in [1,K]$ then $V_{i}^{(T)}$ can be converted as $U_i$, which is a $n \times e$ matrix. 

Every field information in an event log presents its individual significance. To incorporate the relative importance of a word in a particular document we use the term frequency-inverse document frequency (TF-IDF) weighting method\cite{salton1988term}. After applying TF-IDF, the weight vector of an event word vector $V_{i}= [v_1, v_2,\cdots, v_n]$ ($i\in[1,K]$) is represented as $W_{i}=[w_1, w_2, \cdots,w_n]$, where $w_i$ is the TF-IDF score of $v_i$. The content embedding vector $X_k=[X_1, X_2,\cdots, X_K]$ ($k\in[1,K]$) can be obtained using the following equation.

\begin{equation}
    X_k = WU =[x_1 \: x_2 \: \cdots \: x_e], \quad x_j = \frac{1}{n} \sum_{i\in[1,n]} W_i U_{ij}
\end{equation}

The workflow of node content embedding is presented in Figure~\ref{fig:parsing}. The feature vector of the contents is represented as $x \in \Bbb{R}^{e\times1}$ where $e$ is the embedding dimension. This solves the problem of multidimensional host log embedding. In the next step, LogGNN encodes the network topology between the sampled nodes. To obtain the network embedding for each of the nodes of the input graph we use well known word2vec embedding technique. Word2vec takes the sampled neighbor set of the node $v\in V$ and transforms it to a $e-$dimensional embedding vector $E_{net}(v)$.  

\subsubsection{Embedding Aggregation}
As mentioned in section~\ref{subsubsection:neigh_sampling}, the n-hop neighbor encoding vectors $x$ of each node in the graph are aggregated using the BiLSTM model. The neighbor sampling and aggregation are performed for each node of the graph to obtain embedding for all the nodes.

We use a module to aggregate content embeddings using a sequence-based deep learning model BiLSTM\cite{roy2024graphch}. In section ~\ref{subsubsection:neigh_sampling}, we sampled $K$ neighboring nodes for each graph node $v \in K(v)$ using the RWR strategy. Final aggregation is achieved in two steps. In the first step, we concatenate the network embedding $E_{net}(v)$ and content embedding $E_{cont}(v)$. In the second step, the BiLSTM aggregation model takes the concatenated node embedding $E_v = E_{net} + E_{cont}$ of all the $K$ sampled nodes and combines them to capture deep feature interactions. We implement the aggregation model using BiLSTM. The overall aggregation function can be formulated as follows:
\begin{equation}
    E_v^{ag}=\frac{\sum_{v\in {K(v)}}[\overrightarrow{\text{LSTM}}\{E_v\} \oplus \overleftarrow{\text{LSTM}}\{E_v\}]}{K},
\end{equation}
where the $\oplus$ operator denotes the concatenation of left and right directional LSTM output states. The node embedding generated by LogGNN is the input to the statistical clustering and anomaly detection algorithm presented in Section~\ref{sec:detection}. 

\begin{figure*}[!htbp]
  \centering
  \begin{minipage}[b]{\textwidth}
    \includegraphics[width=\textwidth]{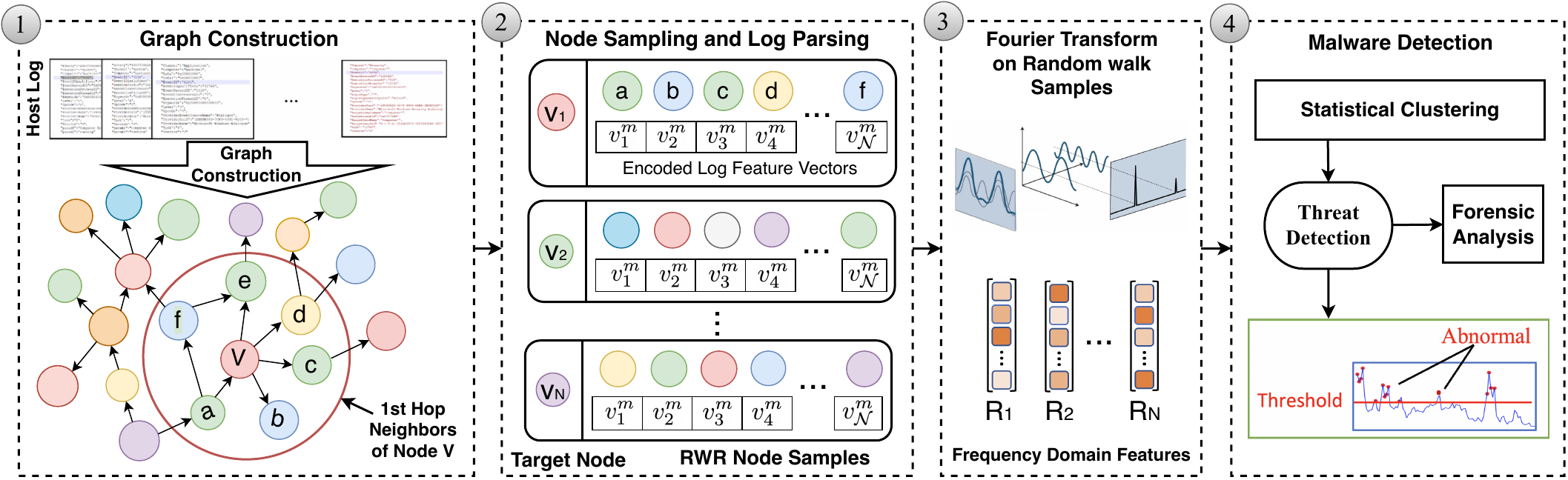}
    \caption{Approach-II: System Architecture of LogSHIELD with frequency domain analysis (FDA).}
    \label{fig:approach_2}
  \end{minipage}
\end{figure*}

\subsection{\textbf{Approach-II:} LogSHIELD with FDA}
\label{subsec:approach_2}

LogSHIELD with FDA uses the same modules of approach-I for graph construction, neighbor sampling, and log parsing as presented in Section~\ref{subsubsec:graph_const} and shown in steps 1 and 2 in Figure~\ref{fig:approach_2}.

In this approach, step 3 performs a frequency domain analysis on the sampled nodes to extract the semantic features of the sampled graph nodes. Let's consider there are N nodes in the graph and we sample $\mathcal{N}(v)$ neighbor nodes of the node $v$. The sampled neighbor set $V$ can be represented as $V = [v_{1}^m, v_{2}^m, \cdots , v_{\mathcal{N}}^m] $, where $m$ is the number of fields in each log.

We perform Discrete Fourier Transformation (DFT) on the $N$ nodes of the graph and $\mathcal{N}$ neighbor samples for each node in N. Considering the $\mathcal{N}$ samples as the window length, we obtain the frequency domain features using the following equation.

\begin{equation}
    F_{ik} = \sum_{n = 1}^\mathcal{N}  x_{kn} e^{-{\frac{2\pi(n-1)(k-1)}{\mathcal{N}}}} \quad (1<k<\mathcal{N}),
\end{equation}
where $k \in [1:N]$ which is $k^{th}$ window frequency component of $\frac{2\pi(k-1)}{\mathcal{N}}$. The frequency domain features $F_{ik} = a_{ik}+jb_{ik}$, are complex numbers and need to transform into real numbers for further clustering.   

\begin{equation}
\begin{cases}
    a_{ik} = \sum_{n = 1}^\mathcal{N} x_{kn}\; cos{ \frac{2\pi(n-1)(k-1)}{\mathcal{N}}} \\
    \\
    b_{ik} = \sum_{n = 1}^\mathcal{N}  -x_{kn} \; sin{ \frac{2\pi(n-1)(k-1)}{\mathcal{N}}}
\end{cases}
\end{equation}

We obtain the modulus of the complex features. As the complex feature vectors from DFT are conjugate and symmetric, we further reduce the feature dimension by taking half of the modulus feature vector. Now taking the first half of the modulus,
\begin{equation}
    r_{ik} = a_{ik}^2+b_{ik}^2
\end{equation}

\begin{equation}
    R_{ik} = \frac{ln(r_{ik}+1)}{C} \quad (1<i<N)
\end{equation}

\begin{equation}
    R_{ik} = R_{ik}[1:L_f] \quad (L_f = \floor*{ \frac{\mathcal{N}}{2}}+1)
\end{equation}

Now we perform a logarithmic transformation on $r_{ik}$ to obtain a numerically stable feature vector. Finally, we obtain the node feature vector $R_{L_f1}$ with dimension $(L \times f1)$.
\begin{equation}
    R_{ki} = 
    \begin{bmatrix}
    R_{11} & R_{12} & \cdots & R_{1N}\\
    R_{21} & R_{22} & \cdots & R_{2N}\\
    \vdots & \vdots & \ddots & \vdots \\
    R_{L_f1} & R_{L_f2} & \cdots & R_{L_f\times N}
    \end{bmatrix},
\end{equation}

where L is the number of log features.
 
\subsection{Anomaly Detection: Statistical Clustering}
\label{sec:detection}
We design an unsupervised statistical clustering algorithm to learn the pattern of log neighbor features in both Approach-I and Approach-II. This is step 4 in both Figure~\ref{fig:architecture} and Figure~\ref{fig:approach_2}. Conventional clustering algorithms like k-means clustering and other similar methods are not suitable because of the initialization of cluster centers and the number of clusters k. Therefore, in the detection phase, we cluster the node embeddings \{$R_1, R_2, R_3, ... R_N$\} using a custom clustering algorithm. We used a cosine-similarity-based log clustering algorithm to cluster system logs and identify malicious activity. We train the clustering algorithm with the benign dataset. Each node embedding $x_i$ from approach-I or feature vector $R_i$ from approach-II where $i\in N$, is assigned to a cluster using their cosine similarity. $C_{n_c}$ clusters are achieved where $n_c$ is the number of clusters. 
 
If $R_N$ is a set of all log feature vectors and $\delta$ is a similarity threshold clustering. It calculates the cosine similarity between two embedding vectors to identify cluster members. The clustering process satisfies the following conditions to obtain clusters \{$C_1, C_2, ... C_{n_c}$\}.

\begin{equation}
  \begin{cases}
    \forall e_x \in C_1, \exists e_y \in C_1,  sim(e_x, e_y) \geq \delta \\
    \forall e_z \in \mathcal{S} \char`\\ C_1 , sim(e_x, e_z) < \delta
  \end{cases}
\end{equation}

where $\mathcal{S} \char`\\ C_1$ is the difference between the two sets. We find the closest cluster center for each embedding or feature vector and calculate the averaged L2-norm representing the training loss.
\begin{equation}
    loss_{train}= \frac{1}{N}\sum_{i=1}^N \| {R_i - \tilde{C}} \|^2
\end{equation}

In the detection phase, the clustering module calculates the distance between the FDA feature of test data sample $R_i^t$ and the benign cluster centers. The closest cluster center is an estimated cluster of $R_i^t$ and calculates the L2-norm as the prediction loss. If the estimation loss is higher than the training loss the log embedding and feature vector are considered malicious. 
\begin{equation}
    loss_{test}= min( \| {R_i - \tilde{C}} \|^2),
\end{equation}
 
where $\tilde{C}$ is the assigned cluster for each vector.

\section{Experimental Evaluation}
\label{sec:experimental_evaluation}
In this section, we evaluate the effectiveness of the LogSHIELD framework and its components in anomaly detection. In addition, we measure the performance of different design parameters. We also perform a thorough ablation study to compare the performance of LogSHIELD with its variants. 

\subsection{Experimental Setup}
LogSHIELD consists of $\sim$8.1k lines of python code and is deployed on a server with Intel(R) Xeon(R) CPU E5-2640 v4 @ 2.40GHz CPU (20 processors) and 48GB memory.

\subsection{\textbf{Dataset}} To evaluate the proposed anomaly detection framework, we use two real-world host datasets. The first dataset is a benign host dataset and the second dataset is a malware dataset with 140 malware sample logs we collected in this work.

\subsubsection{Threat Model}
LogSHIELD endeavors to develop a module for detecting abnormal activities based on host data. We consider only host logs from individual user spaces. As the log resources can be manipulated, the integrity of the host log is out of the scope of this paper and assumes the host logs can be trusted. Due to data limitations, the system environment is limited to Windows OS for benign and malware data samples. The detection system of LogSHIELD is not exposed to any malware data for training and is trained using completely benign data samples. This means that LogSHIELD should be able to detect zero-day attacks. For model evaluation and testing, we collected logs for 140 malware samples of 6 malware families (Ransomware, Rootkit, Trojan, Adware, Backdoor, and Browser Hijack).

\subsubsection{Benign and Malware Dataset}
The benign host log dataset includes system logs from $35$ participants collected over $90$ days. After preprocessing the dataset contains 774M activity logs with a volume of 1.6TB.

The malware dataset contains $374.5K$ logs with a volume of 167MB collected deploying 140 binaries from mostly 6 malware families. Each malware sample was run for 15 minutes, and the logs from the infected machines were stored on a log server. After each run, the infected machines were re-imaged using the FOG Server before proceeding with the next malware sample. The selected and deployed malware samples belong to some of the most disruptive malware families, as shown in Table~\ref{tab:dataset}.

\begin{table}[t]
  \centering
  \caption{Benign and Malware Data Description}
    \begin{tabular}{p{5.5em} p{14em} p{4.385em} p{5.385em} p{7.1em}} \hline
    \textbf{Dataset} & \textbf{Malware Family Count} & \textbf{\#Log} & \textbf{\#Node} & \textbf{Data Volume} \\ \hline
    \textbf{Benign} & x  & 774M & 774M & 1.6TB \\  \hline 
    
    & \textbf{Adware} (3)  & 9K & 9K & 576KB\\  
    & \textbf{Backdoor/Trojan} (45)  & 110K & 110K & 72MB\\ 
    & \textbf{Browser Hijacker} (7)  & 14K & 14K & 4MB\\ 
    \textbf{Malware} & \textbf{Crypto Miner} (5)  & 12.5K & 12.5K & 2.5MB\\ 
    & \textbf{Ransomware} (72)  & 210K & 210K & 86 MB\\ 
    & \textbf{Rootkit} (4)  & 8K & 8K & 2MB\\
    & \textbf{Others} (6)  & 11K & 11K & 670KB\\ \hline
    \end{tabular}%
  \label{tab:dataset}%
\end{table}%

\subsection{Train and Test Datasets}
LogSHIELD is trained in an unsupervised manner with subsets of the benign dataset. The LogGNN embedding model of LogSHIELD approach-I is trained with 35 users' 7 days of benign data (60M of Benign Logs). The detection models of both approach-I and approach-II are trained with 90 days of benign data with 774M of logs. 

To represent the real scenario where malware activities coincide with benign host activities, both approaches of LogSHIELD are evaluated on a malware dataset having around 50\% Benign logs (370K). Besides the 50\% benign data, each malware log with (a 15-minute execution) also contains some benign logs. 


\begin{table}[!htpb]
\small
\caption{Datasets for Training and Testing.}
\label{table:train_test_data}
\centering
\begin{tabular}{p{10.5em} p{12.5em} p{6.5em}}
\hline
\textbf{Models}  &  \textbf{Dataset}  & \textbf{\#Log}\\ \hline
\textbf{LogGNN}   &  \textbf{Train:} Benign   & 60M \\ 
\textbf{Detection Model}   &  \textbf{Train:} Benign   & 774M \\
\textbf{LogSHIELD}   &  \textbf{Test:} Malware & 370K + 372K \\
(Approach - I \& II) & (Benign + Malware Logs)  \\ \hline
\end{tabular}
\end{table}
\subsection{\textbf{LogSHIELD Implementation Details}}
\label{subsec:loggnn_imp}
The components of the LogSHIELD framework both approach-I and approach-II, including log pre-processing and parsing, graph embedding model LogGNN, FDA feature extraction, and downstream clustering models are implemented in Pytorch\cite{pytorch} and the source code will be available soon. 

\indent \textbf{\\Hop Count and Sample Size:} Node sampling in the LogGNN graph embedding model is implemented with $3-hop$ Random Walk with Restart(RWR) and sampled neighbor size is selected K = 40 based on parameter tuning experiments presented in Section~\ref{subsection:parameter_performance} to obtain the optimal semantics of the neighboring nodes.
    
\noindent \textbf{Token Numbers:} In the word2vec log encoding model we set a token size of $5$ from the most frequently seen fields out of a total of $1,200$ unique fields of the training data to obtain the parameter value vectors. 

\noindent \textbf{Word Embedding Dimension:} The embedding dimension is set to $e = 100$ which is determined based on a separate experiment performed with different embedding dimensions presented in Section~\ref{subsection:parameter_performance}. The value of $e$ is passed to the FastText model to encode word parameter value vectors to semantic information vectors. 

\noindent \textbf{BiLSTM Model Deployment:} We use the open-source machine learning library Pytorch to build the BiLSTM model. It consists of 2 hidden layers each layer having 64 and 32 units. The input dimension is 100. The ReLU activation function and the mean absolute error (MAE) loss function is used to train the embedding model for 100 epochs. All the model training and testing were performed using an NVIDIA A100 GPU server with 40GB of memory. We set the minibatch size to $128$, and the learning rate to $0.0015$.

\noindent \textbf{Similarity Threshold:} The similarity threshold $\delta$ between two node embedding or FDA feature vectors is set to $0.72$ for optimal clustering performance. A value of $\delta$ between $0$ to $0.5$ outputs a large number of clusters whereas $\delta$ between $0.8$ to $1$ gives a small number of clusters missing many suspicious clusters.     

\noindent \textbf{DFT Window Size:} In the FDA module the sliding window size of the DFT is set to the number of RWR sampled neighbors $\mathcal{N}=40$. 

\subsection{LogSHIELD Detection Performance}
We perform an anomaly detection experiment to detect malware logs that are never exposed during training. The detection models are trained with benign data only as summarized in Table~\ref{table:train_test_data}. We evaluate the detection performance in terms of Accuracy, Precision, Recall, and F-Score. We also use TPR, FPR, and the area under the ROC curve (AUC) to measure LogSHIELD performance. 

LogSHIELD anomaly detection performance is evaluated across 6 malware families, including Ransomware, Rootkit, Trojan, Adware, Backdoor, and Browser Hijacker. LogSHIELD demonstrated outstanding detection performance for most malware families. Approach-I using LogGNN embedding, achieved an average AUC of 98\%, a TPR of 0.98, and an FPR of 0.038, while Approach-II showed a 97\% AUC, a 0.97 TPR, and a 0.079 FPR, as detailed in Table~\ref{table:detection_result}. However, LogSHIELD's performance with LogGNN marginally degraded for Rootkit and Adware, with AUCs of 96\% and 97\%, respectively. In terms of FPR Approach-I shows 2 times better performance than approach-II with FDA. It confirms that LogGNN graph representation learning is more effective than FDA graph feature learning for detection. This is because LogSHIELD with LogGNN better captures the semantic information of the neighboring nodes compared to LogSHIELD with FDA.    

\begin{table}[!htbp]
  \small
  \centering
  \caption{Malware Detection Results of LogSHIELD [Approach-I: LogSHIELD with LogGNN; Approach-II: LogSHIELD with FDA].}
    \begin{tabular}{p{10.2em} c c c c c c}
    \hline
    \textbf{Malware Families} & \multicolumn{6}{c}{\textbf{Methods}} \\ 
    \multicolumn{1}{c}{} & \multicolumn{3}{c}{\textbf{Approach-I}} & \multicolumn{3}{c}{\textbf{Approach-II}} \\
    \multicolumn{1}{c}{} & \textbf{TPR} & \textbf{FPR} & \textbf{AUC} & \textbf{TPR} & \textbf{FPR} & \textbf{AUC} \\
    \hline 
    \multicolumn{1}{c}{Ransomware} & 0.99 & 0.02 & 0.99 & 0.97 & 0.06 & 0.98 \\
    \multicolumn{1}{c}{Rootkit} & 0.97 & 0.08 & 0.96 & 0.95 & 0.12 & 0.95 \\
    \multicolumn{1}{c}{Trojan} & 0.99 & 0.03 & 0.99 & 0.97 & 0.1 & 0.95 \\
    \multicolumn{1}{c}{Adware} & 0.97 & 0.07 & 0.96 & 0.98 & 0.085 & 0.96 \\
    \multicolumn{1}{c}{Backdoor} & 0.98 & 0.02 & 0.98 & 0.98 & 0.07 & 0.97 \\
    \multicolumn{1}{c}{Browser Hijack} & 0.98 & 0.012 & 0.99 & 0.96 & 0.04 & 0.98 \\
    \hline
    \multicolumn{1}{c}{Average} & 0.98 &	0.038	& 0.98 &	0.97	& 0.079 &	0.97 \\
    \hline
    \end{tabular}%
  \label{table:detection_result}%
\end{table}%


\subsection{LogSHIELD Performance Comparison with Baselines}
We performed another study to evaluate LogSHIELD detection AUC, detection time, and throughput in comparison with the other five baseline anomaly detection models (Log2vec, DeepLog, Unicorn, LogRobust, and DeepRan). According to results presented in Table~\ref{table:class_model_compare}, LogSHIELD with LogGNN embedding outperforms the baseline models in terms of detection performance and achieves 98\% AUC but the detection time of LogSHIELD with LogGNN is similar to baseline models. Whereas LogSHIELD with FDA achieves an AUC of 96\% which is higher than most of the baseline models. It achieves an anomaly detection time of 0.13 seconds which is significantly less than LogSHIELD with LogGNN and other state-of-the-art models. Frequency domain analysis on the provenance graph of host data in approach-II effectively extracts graph representation with significantly less computation than approach-I. Therefore, approach-II, LogSHIELD with FDA significantly outperforms most of the baseline models and takes a small fraction of the detection time of the other models. As FDA uses DFT frequency analysis for log feature extraction replacing deep graph neural network-based (LogGNN), it has less computational overhead. Therefore, LogSHIELD with FDA can provide real-time detection in an enterprise network.

\begin{table}[!htpb]
\small
\caption{Malware Detection Results of LogSHIELD in Comparison with Five Baseline Methods.}
\label{table:class_model_compare}
\centering
\begin{tabular}{p{12.2em} p{3.2em} p{10.2em} p{10.2em}}
\hline
\textbf{Method}  & \textbf{AUC} & \textbf{Detection Time(s)}& \textbf{Throughput (Gbps)}\\ \hline
Log2Vec   & 0.89 & 1.61 & 0.35 \\ 
DeepLog  & 0.95 & 1.65 & 0.5 \\ 
Unicorn  & 0.92 & 0.57 & 3.2 \\ 
LogRobust  & 0.95 & 1.52 & 0.4 \\ 
DeepRan  & 0.98 & 1.36 & 0.72 \\ 
LogSHIELD with LogGNN   & 0.98 & 1.67 & 0.45 \\
LogSHIELD with FDA   & 0.96 & 0.13 & 5.1 \\\hline
\end{tabular}
\end{table}

\subsection{Detection Latency and Throughput}
We conduct experiments on a fused dataset of malware data and 25\% of the large benign dataset. The model is trained with the remaining 75\% benign data. We measure the anomaly detection time and throughput of the detection model using both approach-I and approach-II. The results are shown in Table~\ref{table:class_model_compare}. LogSHIELD with FDA performs significantly better than LogSHIELD with LogGNN. Approach-I with LogGNN can detect malware events with an average detection time of 1.67 seconds whereas approach-II with FDA can detect with an average detection time of 0.13 seconds. The higher computation overhead of LogGNN is one of the main reasons for the higher detection time of approach-I. In terms of throughput, Approach-II achieves 11.3 times higher throughput than approach-I with LogGNN. We find that LogSHIELD with FDA achieves an average throughput of 5.1 Gbps whereas with the LogGNN graph learning model, it achieves 0.45 Gbps. In summary, the results show that the frequency analysis model slightly suffers from detection AUC but significantly outperforms machine learning-based models in terms of detection time and throughput. The throughput of the baseline models is not calculated and compared.

\subsection{Ablation Analysis}
We perform an ablation study using benign and malware datasets to evaluate different functional modules of LogSHIELD in anomaly detection. In this study, we present the significance of RWR node sampling, log graph representation learning method LogGNN, and FDA in the overall detection framework as shown in Table~\ref{table:ablation}. In different combinations, we obtain seven ablation models. All the ablation models are trained with 35 users' benign data only and evaluated with the malware dataset. We use the same provenance graph constructed in previous experiments for all the models. According to the results, $LogSHIELD^1$ and $LogSHIELD^2$ differ on the neighbor sampling and it shows that $LogSHIELD^1$ significantly outperforms $LogSHIELD^2$ with AUC of $0.98$ and $0.89$ respectively as it has the 3-hop neighbor information contributing to semantic node feature extraction. The latter model has less computation time. $LogSHIELD^1$ and $LogSHIELD^4$ differ in graph representation learning algorithm and both use the same neighbor sampling and clustering algorithm. It appears that $LogSHIELD^1$ has a higher computation time which is 1.67 seconds as it uses LogGNN a BiLSTM-based sequence learning model compared to the $LogSHIELD^4$ with a computation time of 0.13 seconds that uses FDA for high-level features of the neighbor samples. $LogSHIELD^2$ and $LogSHIELD^3$ differ in the clustering algorithm keeping other modules unchanged and the results show that  $LogSHIELD^2$ with custom statistical clustering algorithm performs similarly the $LogSHIELD^3$ with k-means clustering in terms of AUC of $0.89$. Without RWR sampling the clustering algorithms are not contributing to better results. Therefore, in $LogSHIELD^4$ and $LogSHIELD^5$, RWR node sampling and FDA are included and the models differ in the clustering algorithm.  Finally, the other two variants $LogSHIELD^6$ and $LogSHIELD^7$ achieve the lowest detection time but without RWR sampling AUC also drops significantly. Therefore $LogSHIELD^4$ presents a better tradeoff between AUC and detection time.

\begin{table*}[t]
\small
\caption{Performance Comparison of LogSHIELD with Different Embedding and Clustering Algorithms.}
\label{table:ablation}
\centering
\begin{tabular}{p{25.2em} p{4.2em} p{4.2em}}
\hline
\textbf{Methods}  & \textbf{AUC} & \textbf{Time}\\ \hline
$LogSHIELD^\textbf{1}$ (RWR+LogGNN+Statistical Clustering)   & 0.98 & 1.67\\ 
$LogSHIELD^\textbf{2}$ (LogGNN + Statistical Clustering)   & 0.89 & 1.5\\
$LogSHIELD^\textbf{3}$ (LogGNN + k-means clustering) & 0.89 & 1.51\\ 
$LogSHIELD^\textbf{4}$ (RWR + FDA + Statistical Clustering)  & 0.97 & 0.13\\ 
$LogSHIELD^\textbf{5}$ (RWR + FDA + k-means clustering)   & 0.95 & 0.15\\
$LogSHIELD^\textbf{6}$ (FDA + Statistical Clustering)   & 0.81 & 0.11\\
$LogSHIELD^\textbf{7}$ (FDA + k-means clustering)   & 0.79 & 0.11\\\hline
\end{tabular}
\end{table*}
\subsection{LogSHIELD Parameter Performance}
\label{subsection:parameter_performance}
\begin{figure}[!t]
\centering
    \begin{subfigure}[b]{0.48\textwidth}
    \centering
        \includegraphics[width=\textwidth]{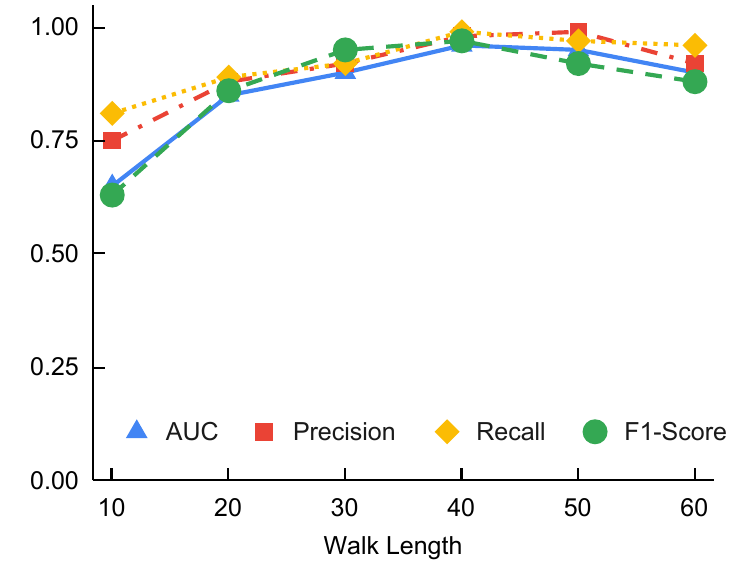}
        \caption{RWR walk length performance measure. Experimented with $\mathcal{N}$ of 10 to 60 nodes in the provenance graph.}
        \label{fig:walk_perf}
    \end{subfigure}%
    \hfill
    \begin{subfigure}[b]{0.48\textwidth}
    \centering
        \includegraphics[width=\textwidth]{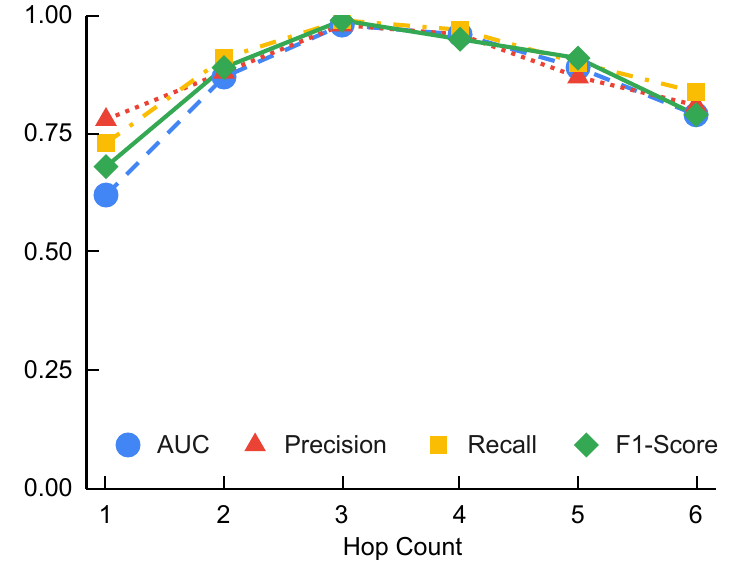}
        \caption{Hop count (n-hop) performance measures keeping walk length fixed. Experimented with hop count $n$ of 1 to 6.}
        \label{fig:hop_perf}
    \end{subfigure}

    \begin{subfigure}[b]{\textwidth}
    \centering
        \includegraphics[width=0.48\textwidth]{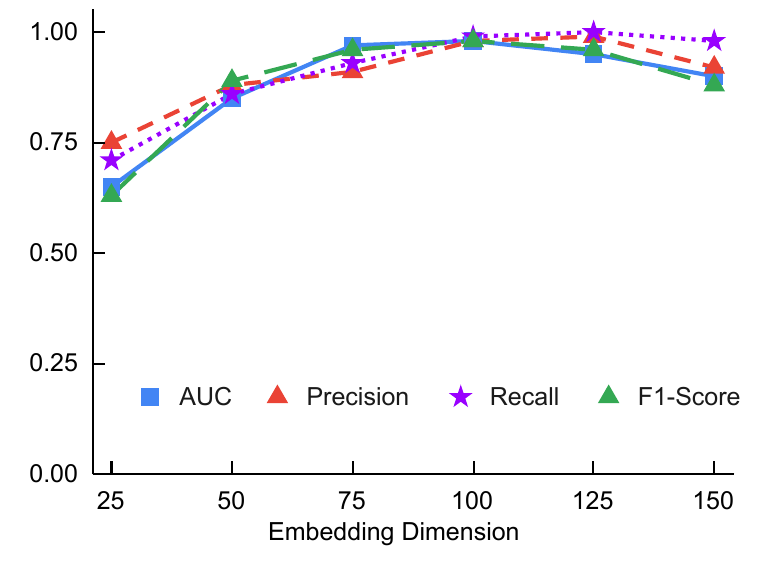} 
        \caption{LogGNN embedding dimension $d$ over performance measures. Experimented with embedding dimensions ranging from 25 to 150.}
        \label{fig:embedding_perf}
    \end{subfigure}

\caption{LogSHIELD performance in terms of AUC, Precision, Recall, and F-Score.}
\label{fig:performance1}

\end{figure}

LogSHIELD performance depends on some key design parameters such as RWR walk length, number of hops for node sampling, LogGNN embedding dimension, etc. To evaluate the significance of the design parameters and find the optimal value we measure the preset performance metrics in different settings. We experiment with random-walk lengths ranging from 10 to 60 with a hop count of 3 as shown in Figure~\ref{fig:walk_perf}. We observe that initially at walk length 10, AUC, Precision, Recall, and F-Score all are very low and gradually increase with higher walk length as more neighbors contribute to the semantic feature of the target node. At a walk length of 40, all the scores reach higher than 0.96 and start gradually decaying at later values. Similar behavior is observed for hop count. We measure performance metrics with hop counts ranging from 1 to 6. At a hop count of 3, LogSHIELD performs with the highest AUC of 0.98 and starts decaying at higher hop counts. The reason for decaying at a higher hop count could be the over-aggregation of unrelated neighbors to target node semantics.

We also measure the model performance with LogGNN embedding dimensions ranging from 25 to 150. The results in Figure~\ref{fig:embedding_perf} show that the model starts to achieve higher results at an embedding dimension of $e = 50$ and gradually increases at higher embedding dimensions at the same time it increases the computation overhead in LogGNN. We chose an embedding dimension of 100 as optimal for LogSHIELD.

\section{Conclusion}
\label{sec:conclusion}
In this paper, we propose a provenance graph-based real-time anomaly detection framework LogSHIELD. We present two graph representation learning approaches using the LogGNN model and FDA frequency analysis for real-time anomaly detection to reduce the computational overhead and improve the efficiency of the detection model. LogSHIELD with LogGNN achieves 98\% average detection AUC whereas with FDA it achieves 96\% AUC. On the other hand, LogSHIELD with FDA takes 0.13 seconds for detection time whereas LogGNN takes 1.67 seconds. It also outperforms baseline models with higher detection AUC and lower detection time. Frequency analysis of host logs with FDA significantly benefits the design of a real-time anomaly detection framework. According to results in Table~\ref{table:ablation}, the novel approach LogSHIELD with RWR and FDA significantly improves anomaly detection efficiency and facilitates real-time detection. A key challenge for LogSHIELD is the preprocessing of high-volume host data and the construction of graphs for detecting advanced persistent threats. In the future, we plan to further analyze the host data to reduce detection time and incorporate more real-host cyber activity. This will allow us to experiment with additional hosts and evaluate the scalability of the framework.

  \section*{Acknowledgments}
This work was supported in part by the U.S. Department of Energy/National Nuclear Security Administration (DOE/NNSA) \#DE-NA0003985, NSF Grants \#1812599, and Sandia National Laboratories (Department of Energy \#DENA000- 352- 5/PO2048463). Any opinions, findings, conclusions, or recommendations expressed in this material are those of the authors and do not necessarily reflect the views of any of these funding agencies.

%
%
%

\bibliographystyle{splncs04}
\bibliography{ref}

\begin{thebibliography}{10}
\providecommand{\url}[1]{\texttt{#1}}
\providecommand{\urlprefix}{URL }
\providecommand{\doi}[1]{https://doi.org/#1}

\bibitem{acquesta2019detailed}
Acquesta, E., Chen, G., Adams, S.S., Bryant, R.D., Haas, J.J., Johnson, N.T., Romanowich, P., Roy, K., Shakamuri, M., Smith, M., et~al.: Detailed statistical models of host-based data for detection of malicious activity. Tech. rep., Sandia National Lab.(SNL-NM), Albuquerque, NM (United States) (2019)

\bibitem{church2017word2vec}
Church, K.W.: Word2vec. Natural Language Engineering  \textbf{23}(1),  155--162 (2017)

\bibitem{du2019lifelong}
Du, M., Chen, Z., Liu, C., Oak, R., Song, D.: Lifelong anomaly detection through unlearning. In: Proceedings of the 2019 ACM SIGSAC Conference on Computer and Communications Security. pp. 1283--1297 (2019)

\bibitem{fu2021realtime}
Fu, C., Li, Q., Shen, M., Xu, K.: Realtime robust malicious traffic detection via frequency domain analysis. In: Proceedings of the 2021 ACM SIGSAC Conference on Computer and Communications Security. pp. 3431--3446 (2021)

\bibitem{gamachchi2017insider}
Gamachchi, A., Boztas, S.: Insider threat detection through attributed graph clustering. In: 2017 IEEE Trustcom/BigDataSE/ICESS. pp. 112--119. IEEE (2017)

\bibitem{gamachchi2018graph}
Gamachchi, A., Sun, L., Boztas, S.: A graph based framework for malicious insider threat detection. arXiv preprint arXiv:1809.00141  (2018)

\bibitem{han2022kernel}
Han, S.H., Lee, D.: Kernel-based real-time file access monitoring structure for detecting malware activity. Electronics  \textbf{11}(12), ~1871 (2022)

\bibitem{han2020unicorn}
Han, X., Pasquier, T., Bates, A., Mickens, J., Seltzer, M.: Unicorn: Runtime provenance-based detector for advanced persistent threats. arXiv preprint arXiv:2001.01525  (2020)

\bibitem{irshad2021trace}
Irshad, H., Ciocarlie, G., Gehani, A., Yegneswaran, V., Lee, K.H., Patel, J., Jha, S., Kwon, Y., Xu, D., Zhang, X.: Trace: Enterprise-wide provenance tracking for real-time apt detection. IEEE Transactions on Information Forensics and Security  \textbf{16},  4363--4376 (2021)

\bibitem{joulin2016fasttext}
Joulin, A., Grave, E., Bojanowski, P., Douze, M., J{\'e}gou, H., Mikolov, T.: Fasttext. zip: Compressing text classification models. arXiv preprint arXiv:1612.03651  (2016)

\bibitem{le2021training}
Le, D.C., Zincir-Heywood, N., Heywood, M.: Training regime influences to semi-supervised learning for insider threat detection. In: 2021 IEEE Security and Privacy Workshops (SPW). pp. 13--18. IEEE (2021)

\bibitem{liu2019log2vec}
Liu, F., Wen, Y., Zhang, D., Jiang, X., Xing, X., Meng, D.: Log2vec: A heterogeneous graph embedding based approach for detecting cyber threats within enterprise. In: Proceedings of the 2019 ACM SIGSAC Conference on Computer and Communications Security. pp. 1777--1794 (2019)

\bibitem{lo2021graphsage}
Lo, W.W., Layeghy, S., Sarhan, M., Gallagher, M., Portmann, M.: E-graphsage: A graph neural network based intrusion detection system. arXiv preprint arXiv:2103.16329  (2021)

\bibitem{panda2022host}
Panda, B., Tripathy, S.N.: Host-specific outlier detection using process relation semantics with graph mining. In: Advances in Data Science and Management, pp. 449--462. Springer (2022)

\bibitem{pytorch}
PRODUCTIO, P.F.R.T.:  (2022), \url{https://pytorch.org/}

\bibitem{roy2024graphch}
Roy, K.C., Chen, G.: Graphch: A deep framework for assessing cyber-human aspects in insider threat detection. IEEE Transactions on Dependable and Secure Computing  (2024)

\bibitem{roy2020deepran}
Roy, K.C., Chen, Q.: Deepran: Attention-based bilstm and crf for ransomware early detection and classifcation. Information Systems Frontiers pp. 1--17 (2020)

\bibitem{salton1988term}
Salton, G., Buckley, C.: Term-weighting approaches in automatic text retrieval. Information processing \& management  \textbf{24}(5),  513--523 (1988)

\bibitem{wang2018learning}
Wang, J., Aggarwal, S., Ji, F., Tay, W.P., et~al.: Learning correlation graph and anomalous employee behavior for insider threat detection. In: 2018 21st International Conference on Information Fusion (FUSION). pp.~1--7. IEEE (2018)

\bibitem{wang2019heterogeneous}
Wang, X., Ji, H., Shi, C., Wang, B., Ye, Y., Cui, P., Yu, P.S.: Heterogeneous graph attention network. In: The World Wide Web Conference. pp. 2022--2032 (2019)

\bibitem{wang2022lightlog}
Wang, Z., Tian, J., Fang, H., Chen, L., Qin, J.: Lightlog: A lightweight temporal convolutional network for log anomaly detection on the edge. Computer Networks  \textbf{203},  108616 (2022)

\bibitem{wu2022paradise}
Wu, Y., Xie, Y., Liao, X., Zhou, P., Feng, D., Wu, L., Li, X., Wildani, A., Long, D.: Paradise: real-time, generalized, and distributed provenance-based intrusion detection. IEEE Transactions on Dependable and Secure Computing  (2022)

\bibitem{yang2022enhanced}
Yang, X., Peng, G., Zhang, D., Lv, Y.: An enhanced intrusion detection system for iot networks based on deep learning and knowledge graph. Security and Communication Networks  \textbf{2022} (2022)

\bibitem{zhang2019heterogeneous}
Zhang, C., Song, D., Huang, C., Swami, A., Chawla, N.V.: Heterogeneous graph neural network. In: Proceedings of the 25th ACM SIGKDD International Conference on Knowledge Discovery \& Data Mining. pp. 793--803 (2019)

\bibitem{zhang2019robust}
Zhang, X., Xu, Y., Lin, Q., Qiao, B., Zhang, H., Dang, Y., Xie, C., Yang, X., Cheng, Q., Li, Z., et~al.: Robust log-based anomaly detection on unstable log data. In: Proceedings of the 2019 27th ACM Joint Meeting on European Software Engineering Conference and Symposium on the Foundations of Software Engineering. pp. 807--817 (2019)

\bibitem{zheng2018realtime}
Zheng, J., Li, Q., Gu, G., Cao, J., Yau, D.K., Wu, J.: Realtime ddos defense using cots sdn switches via adaptive correlation analysis. IEEE Transactions on Information Forensics and Security  \textbf{13}(7),  1838--1853 (2018)

\bibitem{zhu2020you}
Zhu, S., Li, S., Wang, Z., Chen, X., Qian, Z., Krishnamurthy, S.V., Chan, K.S., Swami, A.: You do (not) belong here: detecting dpi evasion attacks with context learning. In: Proceedings of the 16th International Conference on emerging Networking EXperiments and Technologies. pp. 183--197 (2020)

\end{thebibliography}

\end{document}